\documentclass[reprint, superscriptaddress]{revtex4-2}
\usepackage{physics}
\usepackage{amssymb}
\usepackage{algorithm} 
\usepackage{algpseudocode} 
\usepackage{color}
\usepackage{soul}
\usepackage{graphicx}
\usepackage[hyperfootnotes=true, bookmarks=true]{hyperref}
\hypersetup{colorlinks={true},linkcolor={blue},citecolor=blue}
\usepackage[capitalise, nameinlink]{cleveref}

\newcommand{\dif}{\mathrm{d}}
\newcommand{\parentesi}[1]{\left(#1\right)}
\newcommand{\claudator}[1]{\left[#1\right]}
\newcommand{\der}[2]{\frac{\dif #1}{\dif #2}}
\newcommand{\avg}[1]{\left<#1\right>}

\begin{document}

\title{Spatial effects in parasite induced marine diseases of immobile hosts}

\author{\`Alex Gim\'enez-Romero}
\affiliation{Instituto de F\'{\i}sica Interdisciplinar y Sistemas Complejos IFISC (UIB-CSIC), 07122 Palma de Mallorca, Spain}

\author{Federico Vazquez}
\affiliation{Instituto de Cálculo, FCEyN, Universidad de Buenos Aires and CONICET, Buenos Aires, Argentina}
\affiliation{Instituto de F\'{\i}sica Interdisciplinar y Sistemas Complejos IFISC (UIB-CSIC), 07122 Palma de Mallorca, Spain}

\author{Cristóbal López}
\affiliation{Instituto de F\'{\i}sica Interdisciplinar y Sistemas Complejos IFISC (UIB-CSIC), 07122 Palma de Mallorca, Spain}

\author{Manuel A. Mat\'{\i}as}
\affiliation{Instituto de F\'{\i}sica Interdisciplinar y Sistemas Complejos IFISC (UIB-CSIC), 07122 Palma de Mallorca, Spain}

\begin{abstract}
Emerging marine infectious diseases pose a substantial threat to marine ecosystems and the conservation of their biodiversity. Compartmental models of epidemic transmission in marine sessile organisms, available only recently, are based on non-spatial descriptions in which space is homogenised and parasite mobility is not explicitly accounted for.
However, in realistic scenarios epidemic transmission is conditioned by the spatial distribution of hosts and the parasites mobility patterns, calling for a explicit description of space. In this work we develop a spatially-explicit individual-based model to study disease transmission by waterborne parasites in sessile marine populations. We investigate the impact of spatial disease transmission through extensive numerical simulations and theoretical analysis.
Specifically, the effects of parasite mobility into the epidemic threshold and the temporal progression of the epidemic are assessed. We show that larger values of pathogen mobility
imply more severe epidemics, as the number of infections increases, and shorter time-scales to extinction. An analytical expression for the basic reproduction number of the spatial model, $\tilde{R}_0$, is derived as function of the non-spatial counterpart, $R_0$, which characterises a transition between a disease-free and a propagation phase, in which the disease propagates over a large fraction of the system.
\end{abstract}

\maketitle

\section{Introduction}

Wildlife emergent infectious diseases represent a substantial threat to ecosystems and the conservation of their biodiversity \cite{Daszak443}. Their effects can be devastating at the ecological level, causing local extinctions \cite{Daszak443} and in some cases pushing endemic species to the verge of extinction, as is the case of \textit{Pinna nobilis} 
 \cite{Cabanellas2019}; at the economic level, producing losses in agriculture, livestock and aquaculture \cite{Vurro2010, Tomley2009, Pernet2016}, and impact human health, as is the case of the COVID-19 pandemic \cite{Salata2020}. For the past decades, parasites have been continuously emerging \cite{Morens2004, Daszak2017}, while globalisation and climate change have contributed to their evolution. This has allowed these parasites to enter in new ecological niches and spread further the diseases they produce \cite{Aguirre2008}. In particular, marine infectious diseases are recently increasing due to these and other anthropogenic pressures, like pollution and overfishing \cite{Lafferty2004}, inducing widespread mass mortalities in several species \cite{Eisenlord2016, JONES201648, VAZQUEZ2017}.

An important subset of marine organisms affected by infectious diseases are sessile (i.e. they cannot move), like bivalves, sponges or corals. An increasing number of outbreaks affecting marine mollusks have been reported, some of them causing mass mortalities in commercially important bivalves \cite{Guo2016}. Mainly due to the economic importance of some species (e.g. oysters), infectious diseases in bivalve populations have been deeply studied \cite{Petton2021, Pernet2018, McLaughlin2005, powell1999modeling}. Recently, deterministic compartmental models have been used to describe parasite transmitted diseases in marine sessile bivalves \cite{BIDEGAIN_2016_2, BIDEGAIN_perkinsus, GimenezRomero2021}, showing to be able to accurately predict disease transmission in some circumstances. The main limitation of these compartmental models is the assumption of a non-spatial description of the system under study. This underlying hypothesis assumes that any pair parasite-host of the system can interact at any time, which is unrealistic in general. A non-spatial description assumes well mixed populations, which implies that the mean distance among hosts is smaller than the the typical distance explored by parasites in their lifetime. This assumption can be quite realistic in some situations, as it is in \cite{GimenezRomero2021} where the hosts were kept in tanks with water renovation. However, a non-spatial model is not expected to yield a good description of spatially extended hosts in a natural setting.

The key quantity in mathematical epidemiology is the basic reproductive number, $R_0$, that represents the number of infected individuals generated in one generation by the appearance of a single infected individual in a fully susceptible population. Thus, $R_0>1$ ensures the onset of an epidemic, as the number of infected individuals will grow exponentially  producing a widespread disease \cite{Anderson1991}. If we first disregard spatial effects and assume a non-spatial description, $R_0$ can be obtained from standard methods, like the Next Generation Matrix method \cite{Diekmann2010}, and will only depend on \textit{intrinsic} characteristics of the pathosystem (host-pathogen system) under study. However, this basic reproduction number is unable to characterise the threshold behaviour in many situations, including spatially extended systems \cite{Cross2007, Li2011, RILEY201568}. In these systems, the propagation of an epidemic to the entire system needs that a certain spatial threshold is exceeded \cite{Gilligan2008}. Otherwise the disease will only take place in suitable localised parts of the system, not being able to propagate to the total system. Thus, disease spread will be strongly affected by the host spatial distribution and pathogen mobility, which are not accounted for in non-spatial models.

In this work we will try to unravel the transmission mechanisms of a parasite-induced disease affecting immobile hosts in a spatially extended system. We will approach the problem both theoretically and through numerical simulation. The numerical study is based on Individual-Based Modelling (IBM), a method widely used to study ecological systems \cite{Grimm2005}, so that individuals are treated as discrete entities, space is introduced explicitly and the dynamics are stochastic. Representative average behaviours can be obtained by averaging over a sufficient number of realisations, and the accuracy of the approach can be calibrated by deriving the corresponding non-spatial limit, that can be confronted with the suitable compartmental model on which a particular IBM is based. The IBM approach to our problem will allow to study in depth the relation between pathogen mobility and immobile host infection. As parasites move randomly over the space, tracking the position of each parasite at different times turns to be of fundamental importance to properly capture the stochastic dynamics of infections from parasites to hosts. Modelling parasites and hosts as individual entities allows to take into account the spatial and temporal heterogeneity of interactions between them. This heterogeneity and the level of control in microscopic interactions cannot be captured by other mathematical approaches such as partial differential equations. On the other hand, IBMs are mathematically involved, and analytical treatments are normally cumbersome, while their numerical implementation is computationally expensive \cite{Breckling1900}. 

Here we introduce a spatially-explicit individual-based model to study parasite-induced marine diseases of immobile hosts. The model is applied  to the case of diffusing parasites and uniformly distributed hosts. The system under study is an extension of the compartmental model presented in \cite{GimenezRomero2021} and, to our best knowledge, it is the first stochastic spatially-explicit model for the study of marine epidemics with mobile parasites and immobile hosts. As a main result, we find that the occurrence of an outbreak will depend on the balance between the intrinsic characteristics of the pathosystem, well represented by the above described non-spatial basic reproductive number, $R_0$, and features that characterise parasite mobility. We generalise the basic reproductive number, that we will refer to as $\tilde{R}_0$, such that it accounts for the number of hosts that get infected by the appearance of a single infected individual in a fully susceptible population in a spatially extended system. $\tilde{R}_0$ characterises the global epidemic and can be written as a product between $R_0$ and a factor describing parasite mobility. The latter factor is smaller and at most equal to $1$, which implies that, as it could be expected, it is more difficult to induce a global outbreak in a spatially extended system (a two dimensional lattice in our case) that in a well mixed (non-spatial) population. 

The paper is organised as follows: in \cref{sec:model}, we introduce some biological considerations for bivalve epidemics, discussed in more detailed in Ref. \cite{GimenezRomero2021}, and build the spatially-explicit model. In \cref{sec:results}, we present analytical results that are discussed and supported by numerical simulations. Specifically, the high mobility limit is discussed and connected to the the compartmental model. An approximation for the parasite population is discussed. Then, the effect of parasite mobility to the epidemic threshold is characterised, deriving an analytical expression for the basic reproduction number. Furthermore, the spreading speed of the disease and the time-scale to extinction is investigated. Finally, \cref{sec: conclusions} contains some concluding remarks.

\section{The SIRP spatial model} \label{sec:model}

The most important biological features of the system under study are as follows. First, hosts are immobile, while the disease is transmitted by parasites produced by infected hosts. There are two mechanisms by which parasites are cleared from the medium: i) they have a finite life time after which they die; ii) they get absorbed after they infect a host and thus are no longer in the medium and cannot infect other hosts. Recruiting (birth) of hosts occur at a very slow rate compared to other timescales in the system, and accordingly it will be considered negligible in the model. Moreover, hosts do not show long-term immunity, as is typical of invertebrates, like mollusks \cite{Powell2015}. We also assume that recovery (healing) of infected hosts, if it occurs, can be neglected. Furthermore, we consider that dead hosts are not a source of parasites in the medium. See \cite{GimenezRomero2021} for a detailed presentation of the non-spatial SIRP model, including these biological modelling considerations.

Under these considerations, we introduce an individual-based model with explicit space characterisation to study the effect of parasite mobility in disease transmission. We consider a square grid of length $L$ with periodic boundary conditions and place a single host per site, so that there are $N=L^2$ hosts. The hosts can be in three discrete states: susceptible, $S$; infected, $I$ and dead (or removed), $R$. Then, we introduce the parasite population as a new individual with a single state, $P$. Hosts are sessile (i.e. immobile), while parasites are allowed to move between the lattice sites. As initial condition we assume that the entire host population is susceptible, $S(0)=N=L^2$, and that a small initial number of parasites, $P(0)$ is introduced in the system.

\begin{figure}[H]
    \centering
    \includegraphics[width=0.9\columnwidth]{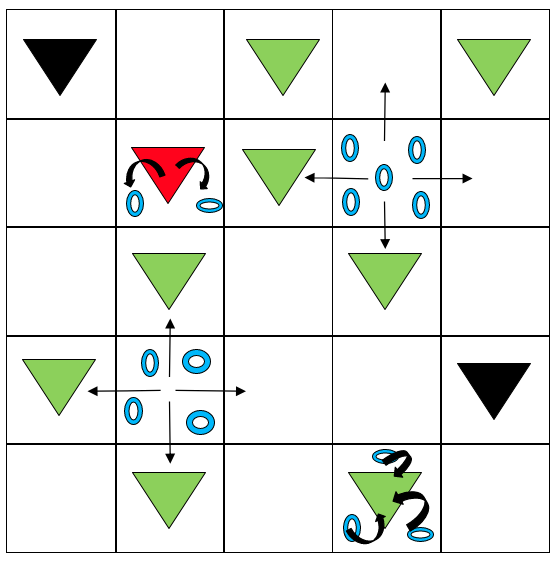}
    \caption{Scheme of the individual based model. Green, red and black triangles represent susceptible, infected and dead hosts, respectively. Blue rings represent parasites, which move randomly between cells. Susceptible hosts get infected by filtering parasites while infected hosts produce them. Dead hosts do not participate in the dynamics of the system.}
    \label{fig:IBM}
\end{figure}

Infection occurs when susceptible hosts filter parasites in their close proximity. Accordingly, the infection process is implemented between parasites and susceptible hosts sharing the same lattice site. In particular, susceptible hosts in contact with a parasite become infected at rate $\beta$. As the infection event implies the filtering of a parasite by a susceptible host, when a new infection occurs a parasite of that particular site is removed. Infected individuals die at rate $\gamma$ and produce parasites at rate $\lambda$, while parasites die at rate $\mu$. Parasites move randomly between the four neighbouring lattice sites at rate $\kappa$, which corresponds to a diffusive motion. \cref{fig:IBM} shows a schematic representation of the dynamics and \cref{eq:scheme} summarises the reactive events
 \begin{equation}\label{eq:scheme}
    S+P \stackrel{\beta}{\rightarrow} I + \varnothing \quad I  \stackrel{\gamma}{\rightarrow} R \quad I \stackrel{\lambda}{\rightarrow} I+P \quad P \stackrel{\mu}{\rightarrow} \varnothing \ .
\end{equation}   

Formally, the model is mathematically described by a system of $N$ master equations for the probabilities of the states in each lattice site $i$. This is very difficult to manage analytically, so the time evolution of the model is numerically solved using Gillespie's algorithm \cite{Gillespie1977} (the code can be found in \cite{CODE}).

\begin{table}[H]
    \centering
    \caption{Variables and parameters of the model}
    \resizebox{\columnwidth}{!}{
    \begin{tabular}{cl}
    \hline \hline
        \textbf{Variable/Parameter} & \textbf{Definition} \\ \hline
        S & Susceptible host \\
        I & Infected host \\
        R & Dead host \\
        P & Parasite \\
        $\beta$ & Parasite-host transmission rate \\
        $\gamma$ & Host mortality rate \\
        $\lambda$ & Production rate of parasites by infected hosts \\
        $\mu$ & Parasite natural death rate \\
        $\kappa$ & Parasite dispersal rate (mobility) \\
        $R_0$ & Non-spatial basic reproductive number \\
        $\tilde{R}_0$ & Spatial basic reproductive number
        \\ \hline \hline
    \end{tabular}
    }
    \label{tab:my_label}
\end{table}
    
\section{Results}\label{sec:results}
In this section several features of the model are studied, both numerically (from IBM simulations) and analytically. All numerical results were obtained for a square lattice of length $L=100$, with $N=S(0)=10^4$ hosts and using a small initial condition of $P(0)=50$ parasites in the centre site. 

\subsection{Non-spatial limit} \label{sec:MFlimit}
An important test of the IBM implementation is to show that, under suitable circumstances, it converges to the non-spatial model on which the IBM is based. This occurs in the limit when the parasites move many times before dying or infecting a susceptible host. In this situation, each parasite visits typically all the hosts of the system and may infect any of them. This is equivalent to infecting a random host of the system, which happens with probability $\beta S/N$, being $S$ the total number of susceptible hosts in the system. An equivalent picture is that parasites will end up uniformly distributed in the lattice, so that there will be $P/N$ parasites in each lattice site at any time. One expects to reach these conditions when $\kappa\gg\mu,\beta$, and, thus the system as a whole can be described by the following system of ordinary differential equations (ODE's), 
\begin{equation}\label{eq:SIRP_MF}
    \begin{aligned}
        \dot{S} &=-\beta P S/N \, , \\
        \dot{I} &=\beta P S/N-\gamma I \, , \\
        \dot{R} &=\gamma I \, , \\
        \dot{P} &=\lambda I-\beta P S/N-\mu P \ ,
    \end{aligned}
\end{equation}  
that is precisely the SIRP non-spatial model \cite{GimenezRomero2021}, where $S$, $I$, $R$ are the total number of susceptible, infected and recovered hosts in the system, $P$ the total number of parasites and $N$ is the number of hosts. 

The basic reproduction number, $R_0$, of this non-spatial model is the dimensionless quantity that yields the number of secondary infections generated by the appearance of a single infected individual in a completely susceptible population, also indicating whether the system will exhibit an epidemic outbreak, $R_0>1$, or not, $R_0<1$.
In our case it can be directly computed as the mean number of parasites produced by an infected host during its mean lifetime, $\lambda/\gamma$, times the mean number of susceptible hosts that get infected by parasites during their mean lifetime, $\beta/(\mu+\beta)$,
\begin{equation}\label{eq:R0_MF}
    R_0=\frac{\lambda}{\gamma}\frac{\beta}{\mu+\beta} \ .
\end{equation}
This result can be corroborated with standard methods such as the Next Generation Matrix method \cite{Diekmann2010} (see \cite{GimenezRomero2021}), where $S(0)=N$ has been considered.

Moreover, the model has a conserved quantity $\mathcal{C}$ \cite{GimenezRomero2021} that allows to find an analytical expression for the final number of dead individuals (cf. \cref{app:Rinf}),
\begin{equation}\label{eq:R_inf}
    R(\infty)=N+\frac{S(0)}{\xi}W_0\parentesi{-\xi\exp(-\frac{\beta}{\mu}C)} \ ,
\end{equation}
with $\displaystyle\xi=S(0)\frac{\beta\parentesi{\lambda-\gamma}}{\mu\gamma}$ and $\displaystyle C=P_0+\frac{\lambda}{\gamma}\parentesi{S(0)+I(0)}-S(0)$.

\begin{figure}[H]
    \centering
    \includegraphics[width=\columnwidth]{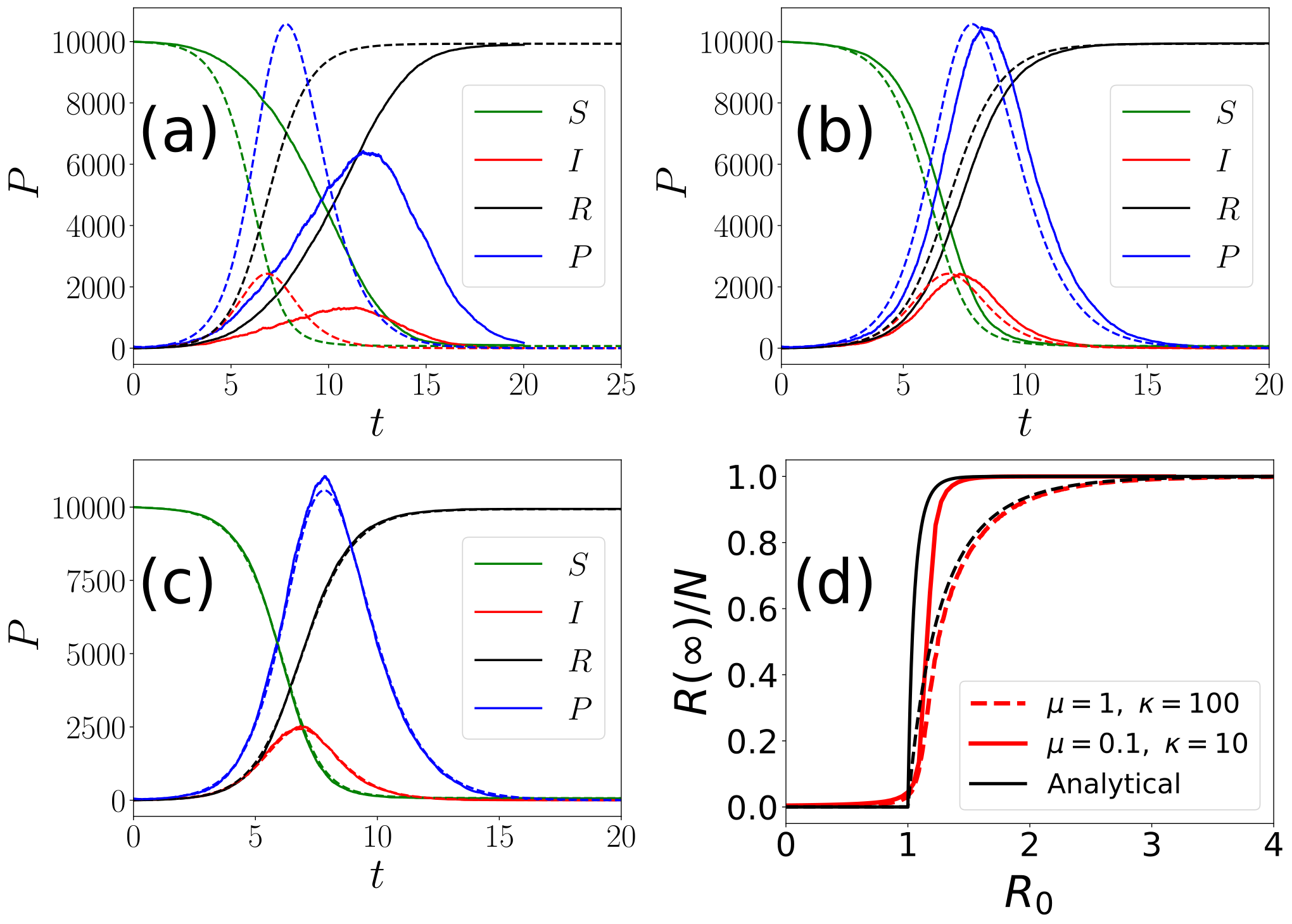}
    \caption{Numerical solution of the non-spatial model (\cref{eq:SIRP_MF}, dashed lines) compared with numerical solutions of the individual based model (solid lines) approaching the non-spatial limit with fixed $\gamma=\mu=\beta=1$ and $\lambda=6$. (a) $\kappa=10^2$, (b) $\kappa=10^3$, (c) $\kappa=10^4$. Panel (d) shows the final fraction of dead hosts, $R(\infty)/N$, as function of $R_0$ for $\kappa/(\mu+\kappa)=0.999$ with $\mu=1,0.1$ compared to the analytical result.}
    \label{fig:MF_limit}
\end{figure}

The non-spatial limit of the model has been evaluated by comparing realisations of the stochastic model (in the limit $\kappa\gg\mu,\beta$) with numerical solutions of the non-spatial ODE system of \cref{eq:SIRP_MF}. Furthermore, the analytical expression for $R(\infty)$ using the non-spatial model, \cref{eq:R_inf}, is also compared to the numerical results of the individual based model. As shown in \cref{fig:MF_limit} (a)-(c), as $\kappa$ is increased compared to $\mu$ the individual based model approaches the non-spatial one. \cref{fig:MF_limit}(d) shows how the numerical results for $R(\infty)$ for different $R_0$ values approach the analytical solution in the non-spatial limit. 

\subsection{Approximate relation between parasites and infected hosts}

    In the limit $\kappa\gg\beta,\mu$ a time-scale approximation can be performed so that the parasite population dynamics directly relates to that of the infected hosts. In the non-spatial limit it was already shown in \cite{GimenezRomero2021} that, if  $\mu\gg\beta,\gamma$ and $\lambda\gg\beta P/N$, the total parasite population of the system can be well described using the approximation (see \cite{GimenezRomero2021} for a detailed discussion),
    \begin{equation}\label{eq:P_approx}
        P(t)\approx \frac{\lambda}{\mu} I(t) \ ,
    \end{equation}
    
    Here we extend the validity of this approximation to spatial systems far from the non-spatial limit. Consider the local dynamics of the parasite population on a lattice site $i$. Note that when the host in the site is susceptible, parasites in this site can either infect the host, die, or move to another site. All these processes imply that a parasite will disappear from the current site. Once the host at site $i$ gets infected, infection can no longer occur whereas parasite production is now possible. If $\kappa$ is small enough compared to $\lambda$ and $\mu$,  the only competing processes in sites with infected individuals will be the production of parasites and their natural death, which can be fairly described by the following rate equation,
    \begin{equation}
        \der{P}{t}=\lambda-\mu P \ ,
    \end{equation}
    whose solution is
    \begin{equation}\label{eq:P_evol}
        P(t)=\frac{\lambda}{\mu}+\claudator{P(0)-\frac{\lambda}{\mu}}e^{-\mu t} \ .
    \end{equation}
    
    From \cref{eq:P_evol} one may notice that the stationary value of $P$, $\lambda/\mu$, is reached in a time proportional to $t_{\textrm{eq}}\propto 1/\mu$. This derivation allows to find a condition for which \cref{eq:P_approx} is valid beyond the non-spatial limit. Basically, if the mean dispersal time, $1/\kappa$, is greater than the equilibrium time, $t_{\textrm{eq}}\propto 1/\mu$, parasites in sites with infected hosts will reach its stationary level before parasites enter or leave the sites. Thus, sites with infected hosts can be considered as a closed system and the approximation holds. In other words, if the dispersal rate of parasites is small compared to the parasite deactivation rate, $\kappa \ll \mu$, the local parasite population of the site will reach its stationary level $P_i=\lambda/\mu$. It is possible to extend the result to the entire system: if there are $I(t)$ infected sites in the system at time $t$ and $\kappa\ll\mu$ is fulfilled, there will be a total parasite population of $P(t)=(\lambda/\mu) I(t)$, which is equivalent to \cref{eq:P_approx}.
    
    Thus, for the non-spatial limit ($\kappa\gg\mu$) we have that if $\mu\gg\beta,\gamma$ \cref{eq:P_approx} is valid, while for $\kappa\ll\mu$ the approximation is also valid regardless of the value of $\beta,\gamma$, as the nature of the approximation is different. Thus, in general, as $\kappa$ decreases over $\mu$ (the lower the parasite mobility becomes) we expect the approximation to work better.
    
\begin{figure}[H]
        \centering
        \includegraphics[width=\columnwidth]{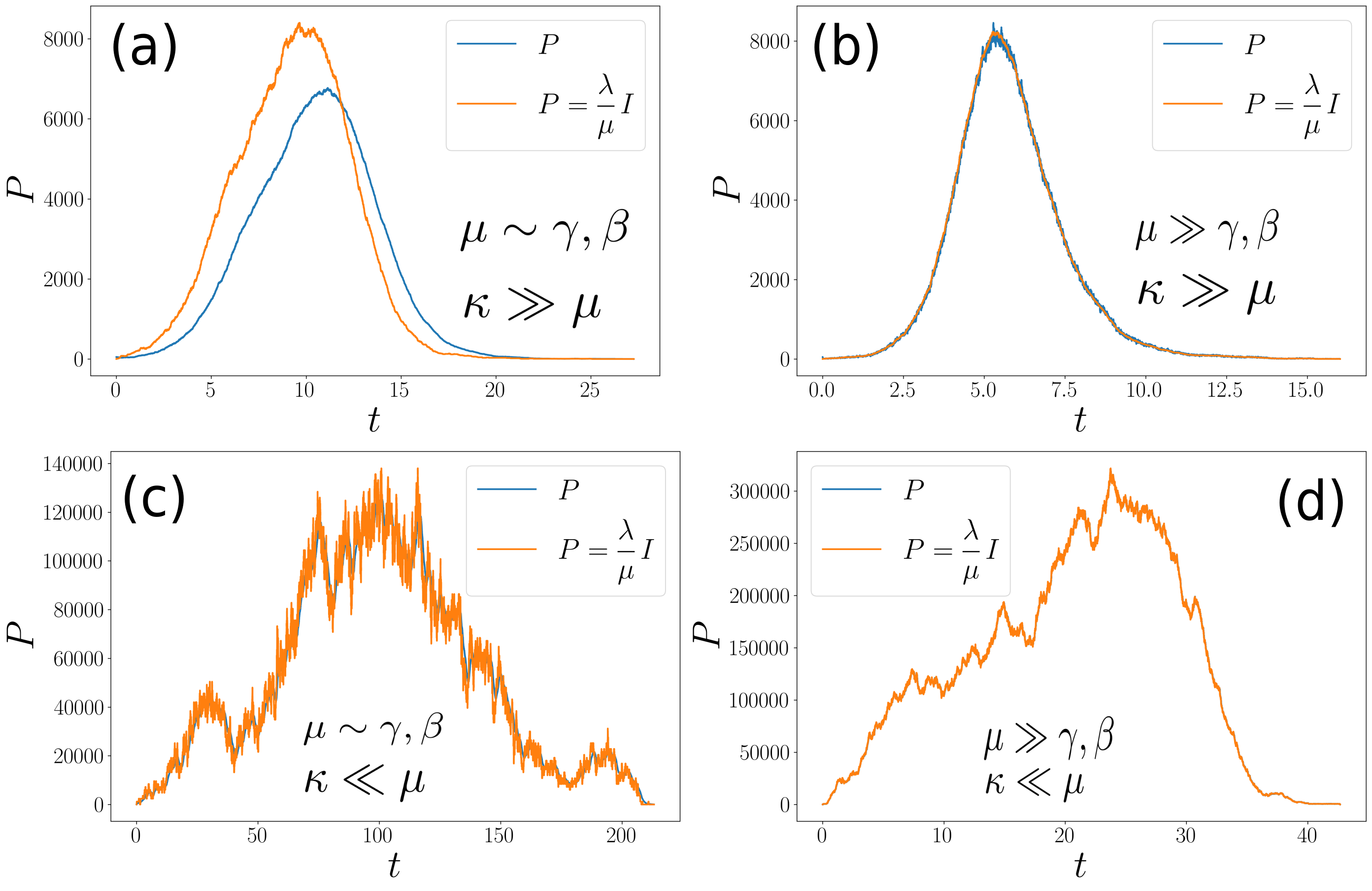}
        \caption{Numerical verification of the approximate expression for the parasite population dynamics, \cref{eq:P_approx}, for different mobility conditions. The simulations were performed fixing $\beta=\gamma=1$  for all panels. (a) $\mu=1$ $\kappa=10^2$, $\lambda=6.06$, $\kappa/(\mu+\kappa)=0.99$; (b) $\mu=100$, $\kappa=10^4$, $\lambda=306$,  $\kappa/(\mu+\kappa)=0.99$; (c) $\mu=1$, $\kappa=0.01$, $\lambda=1200$, $\kappa/(\mu+\kappa)=0.01$; (d) $\mu=100$, $\kappa=1$, $\lambda=60600$,  $\kappa/(\mu+\kappa)=0.01$} 
        \label{fig:P_approx}
    \end{figure}

    The parasite approximation to infected hosts dynamics, \cref{eq:P_approx}, is numerically verified for different mobility conditions. -\cref{fig:P_approx}(a)-(b) shows how the approximation improves as $\mu$ grows over $\beta,\gamma$ (mean errors are $0.18$ and $0.0081$, respectively) in the non-spatial limit, i.e. $\kappa\gg\mu$, as expected. This result is in perfect agreement with that found in \cite{GimenezRomero2021}. Then, \cref{fig:P_approx}(c)-(d) show that the approximation is valid in general when $\kappa\ll\mu$ but improves anyway when $\mu\gg\beta,\gamma$ (mean errors are 0.04 and 0.0026, respectively). Summarising, we see that the lower the value of $\kappa$ is with respect to $\mu$ the more valid \cref{eq:P_approx} is, regardless of the value of $\beta,\gamma$, while in the non-spatial limit, $\kappa\gg\mu$, the condition $\mu\gg\beta,\gamma$ is needed.
    
    \begin{figure*}[t!]
        \centering
        \includegraphics[width=1\textwidth]{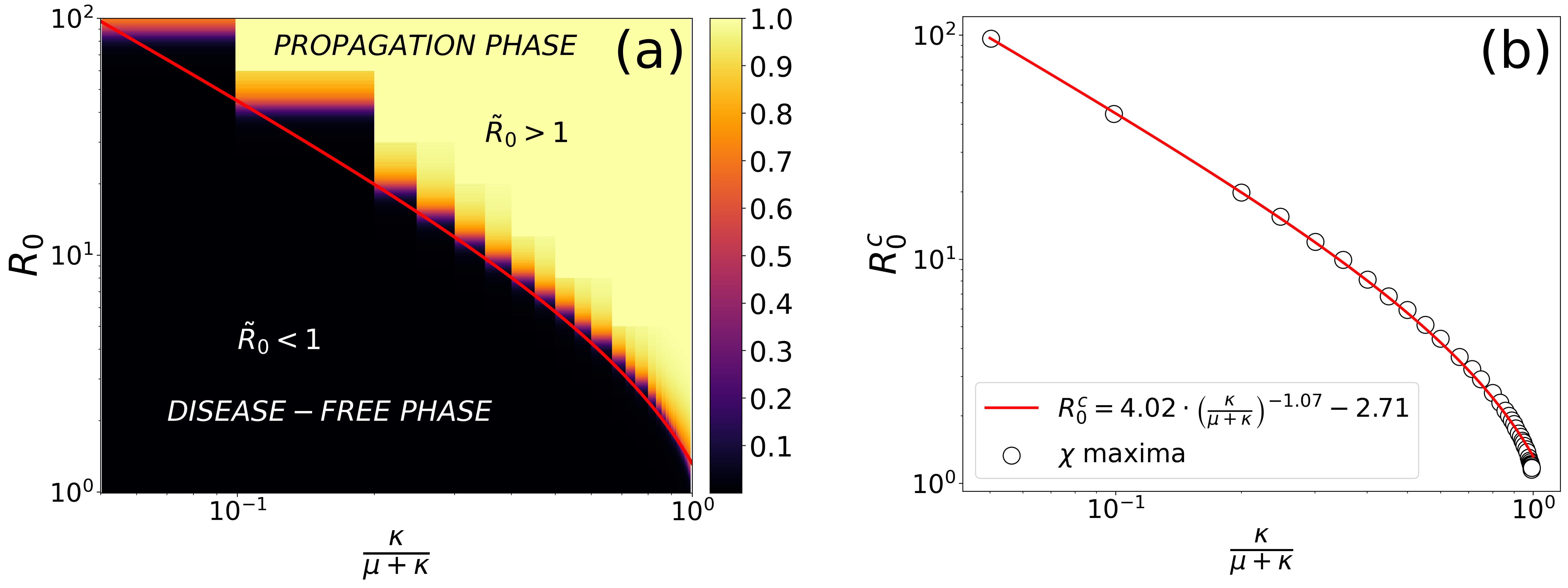}
        \caption{(a) Phase diagram showing the transition between the the disease-free phase and the propagation phase for several values of the parasite mobility and $R_0$. The colour code represents the fraction of dead individuals (i.e. $R/N$) in the final state of the epidemic computed by the average over 1000 realisations. (b) Fit for the transition line following \cref{eq: threshold_corrections}, where dots are the maximums of the ``order parameter'' fluctuations, $\chi=\avg{{R(\infty})^2}-\avg{R(\infty)}^2$}
        \label{fig:Phase_diagram}
    \end{figure*}
    
\subsection{Spatial threshold}

    One of the main questions in epidemiology is to define the conditions under which an epidemic outbreak occurs, which usually is translated into the existence of a threshold. In a well mixed (non-spatial) system the basic reproduction number ($R_0$), that characterises this threshold $R_0=1$, can be defined exclusively from \textit{intrinsic} parameters of the pathosystem, as the host-pathogen interaction does not depend on the host spatial structure or pathogen mobility (see \cref{eq:R0_MF}). In stochastic spatial models this formulation of $R_0$ breaks down. First of all, in stochastic models, even above the threshold there is a non-zero probability that the disease is unable to propagate initially, given by $P_{\textrm{outbreak}}=1-\parentesi{1/R_0}^{I(0)}$ \cite{Brauer2008}. Furthermore, the discrete nature of the populations also modifies the estimates of $R_0$ \cite{KEELING200051}. On the other hand, the introduction of space changes completely the nature of epidemic outbreaks, modifying the host-pathogen interactions by means of specific host spatial distributions and pathogen mobility patterns. Even if the basic reproduction number of the non-spatial model is above the threshold ($R_0>1$), if parasite mobility is not large enough, the epidemic will stay locally confined. Thus, one expects that the threshold at which an epidemic outbreak can propagate to the rest of the system will depend on the balance between the intrinsic pathosystem parameters in $R_0$ and parasite mobility, defining a spatial basic reproduction number, $\tilde{R}_0$.
    
    Having in mind the study in \cref{sec:MFlimit}, we expect that in the high mobility limit the basic reproduction number is defined by the non-spatial formula, \cref{eq:R0_MF}. On the other hand, the lower the parasite mobility is, the more difficult will be for a local outbreak to propagate through the system. Thus, it is natural to think of an spatial basic reproduction number of the form $\tilde{R}_0=R_0 f(\kappa)$, where $f(\kappa)$ is an increasing function of the parasite dispersal rate accounting for parasite mobility fulfilling the limit $\lim_{\kappa\to\infty}f(\kappa)=1$.
    
    Indeed, some authors recently showed that the spatial basic reproduction number can be defined as the product between the non-spatial value, $R_0$, and a factor accounting for spatially-dependent interactions, $f(r)$, in the form $\tilde{R}_0=R_0f(r)$ \cite{Filipe2003, Gilligan2021}. However, these expressions are not analytical \cite{Filipe2003,Filipe2004} or are not directly related to pathogen mobility \cite{Gilligan2021}. Here we propose a simple expression for the spatial basic reproduction number regulating the spatial propagation of the epidemic,
    \begin{equation}\label{eq:R_0ibm}
        \tilde{R}_0=\frac{\lambda}{\gamma}\frac{\beta}{\mu+\beta}\frac{\kappa}{\mu+\kappa}=R_0\frac{\kappa}{\mu+\kappa} \ .
    \end{equation}
    
    The derivation of \cref{eq:R_0ibm} accounts for the number of secondary parasites that are able to produce new infections, or equivalently, the number of secondary infections produced by an initial infected host. If we consider an initial infected individual, on average it will produce $\lambda/\gamma$ parasites. Then, these parasites can only move to neighbouring sites or die, so that the dispersal probability is given by $\kappa/(\mu+\kappa)$. Finally, considering that parasites do not affect each other trying to infect the same host, the infection probability is given by $\beta/(\mu+\beta)$. Joining all terms, we finally obtain \cref{eq:R_0ibm}. This expression is valid when parasites move only to sites with susceptible individuals and do not try to infect the same host. Thus, the derived $\tilde{R}_0$ is only an approximation to the spatial basic reproduction number for the case of an initial introduction of a small quantity of parasites in a fully susceptible population.
    
    Note that, as expected, the spatial basic reproduction number is nothing other than the basic reproduction number of the non-spatial model multiplied by an increasing function of the parasite mobility, $\kappa/(\mu+\kappa).$ Taking the limit $\kappa\gg\mu$ in \cref{eq:R_0ibm} (non-spatial limit) the basic reproduction number of the non-spatial model is recovered. Conversely, in the limit of very low mobility, the $\kappa/(\mu+\kappa)$ factor is small, and this has to be compensated with a large value of the non-spatial basic reproduction number, $R_0$, in order that there is an outbreak, i.e., $\tilde{R}_0>1$.

    The spatial threshold, $\tilde{R_0}=1$, given by \cref{eq:R_0ibm}, has been numerically checked by computing the phase diagram between the absorbing phase $R(\infty)\approx 0$ (no infection, i.e. disease-free state)  and the active phase $R(\infty)> 0$ (in which some level of infection has occurred, i.e. propagation phase) for several values of the parasite mobility and the basic reproduction number of the non-spatial model, $R_0$ \cref{eq:R0_MF}. The transition is expected to occur at $\tilde{R}_0=1$, implying from \cref{eq:R_0ibm} that the dependence of the critical value of $R_0$, say $R_0^c$, is expected to take the form,
    \begin{equation}\label{eq: R_0_crit dependence}
        R_0^c\sim\parentesi{\frac{\kappa}{\mu+\kappa}}^{-1} \ .
    \end{equation}
    As discussed above, we expect that $\tilde{R}_0=1$ with $\tilde{R}_0$ given by \cref{eq:R_0ibm} does not represent exactly the spatial threshold, and for this reason we suggest the more general functional form,
    \begin{equation}\label{eq: threshold_corrections}
        R_0^c\sim A\parentesi{\frac{\kappa}{\mu+\kappa}}^{-B} - C\ ,
    \end{equation}
    to be be fitted to numerical data, where $A=1$, $B=1$ and $C=0$ would imply a perfect agreement of numerical simulations of the IBM model with \cref{eq:R_0ibm}.
    
    In order to obtain the phase diagram, we compute the absorbing state of the model as an average over $1000$ realisations for each value of the mobility and $R_0$ considered. Then, the critical value $R_0^c$ is computed for each mobility value as the $R_0$ value for which the fluctuations of the ``order parameter'' $\displaystyle \chi=\avg{{R(\infty})^2}-\avg{R(\infty)}^2$ are maximal, as this would be an indication of a transition between the disease-free and the propagation phases.

    \cref{fig:Phase_diagram}(a) shows the numerical results of the computed transition between the disease-free and propagation phases. The heatmap coding represents the average value of absorbing state $\avg{R(\infty)}$ for several values of the mobility factor and $R_0$. As expected, the lower the mobility factor is, the higher the value of $R_0$ is needed for the disease to invade the population. \cref{fig:Phase_diagram}(b) shows the fit of \cref{eq: threshold_corrections} with less than a $1\%$ of relative error. Interestingly, we obtain $B=1.07\approx1$ which validates our expression for the spatial threshold as a first approximation. However, the values for $A=4.02$ and $C=2.7$ show a significant deviation from \cref{eq: R_0_crit dependence} and indicate that the expression \cref{eq:R_0ibm} is  an approximation to the spatial basic reproduction number, which however seems to contain the right dependence on $\kappa/(\mu+\kappa)$, and where $A$ could be a geometric factor for a lattice.
    
\subsection{Spreading speed of the infected population and time to extinction}

    Another relevant epidemiological question is how does an infected population spread after the onset of an epidemic. In order to obtain this spreading speed we computed the mean time needed for an infected individual to reach the boundary of the system. More specifically, for each particular choice of the model parameters, 1000 simulations were run for several system sizes ranging from $L=10$ to $L=60$. The computed mean time was found to depend linearly with the system size, thus allowing to compute the speed from the slope of this relation. With this procedure, the spreading speed was computed for several values of the parasite mobility and $R_0$, large enough to ensure an epidemic outbreak that reached the boundary of the system. In this situation, the spreading speed is expected to depend linearly with the square root of the parasite mobility, 
    \begin{equation}\label{eq:front_vel}
        v\sim\sqrt{\kappa} \ .
    \end{equation}
    
\begin{figure}[H]
        \centering
        \includegraphics[width=\columnwidth]{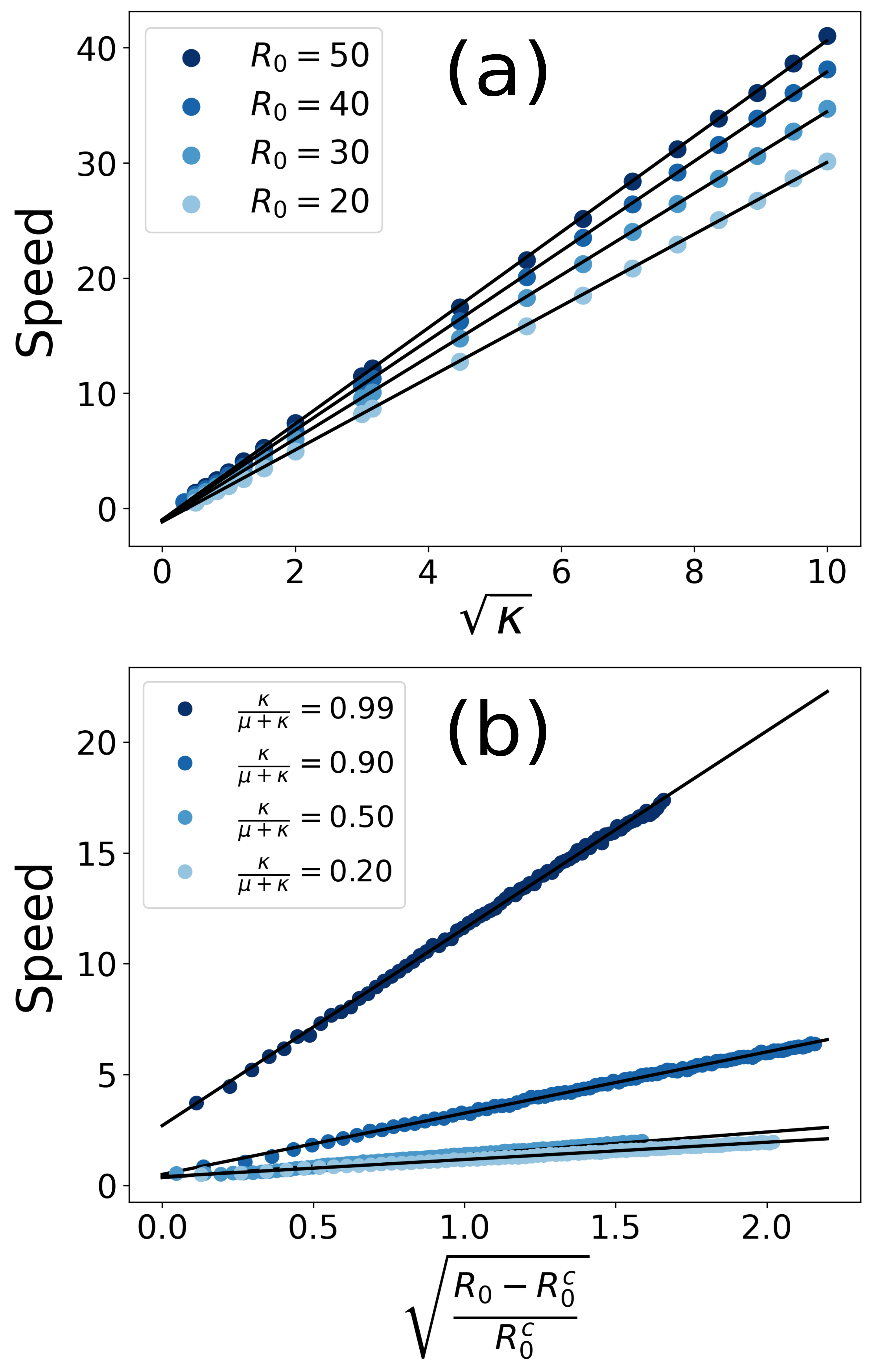}
        \caption{(a) Disease spreading speed as function of the square root of the parasite mobility for several values of $R_0$. The plot shows a remarkable agreement with \cref{eq:front_vel}. (b) Disease spreading speed as function of the square root of the distance to the critical value of $R_0$ for several values of the parasite mobility.}
        \label{fig:front_velocity}
    \end{figure}

    \cref{fig:front_velocity}(a) shows this square root dependence for different values of the fixed $R_0$. Similarly, the speed was also computed for several values of the basic reproduction number and a fixed mobility. In this case, it varies with the square root of the distance to the critical value of $R_0$, $R_0^c$, as shown by \cref{fig:front_velocity}(b). This is in good agreement with other mathematically similar models \cite{Bertuzzo2010}.
    
    \begin{figure*}[t!]
        \centering
        \includegraphics[width=\textwidth]{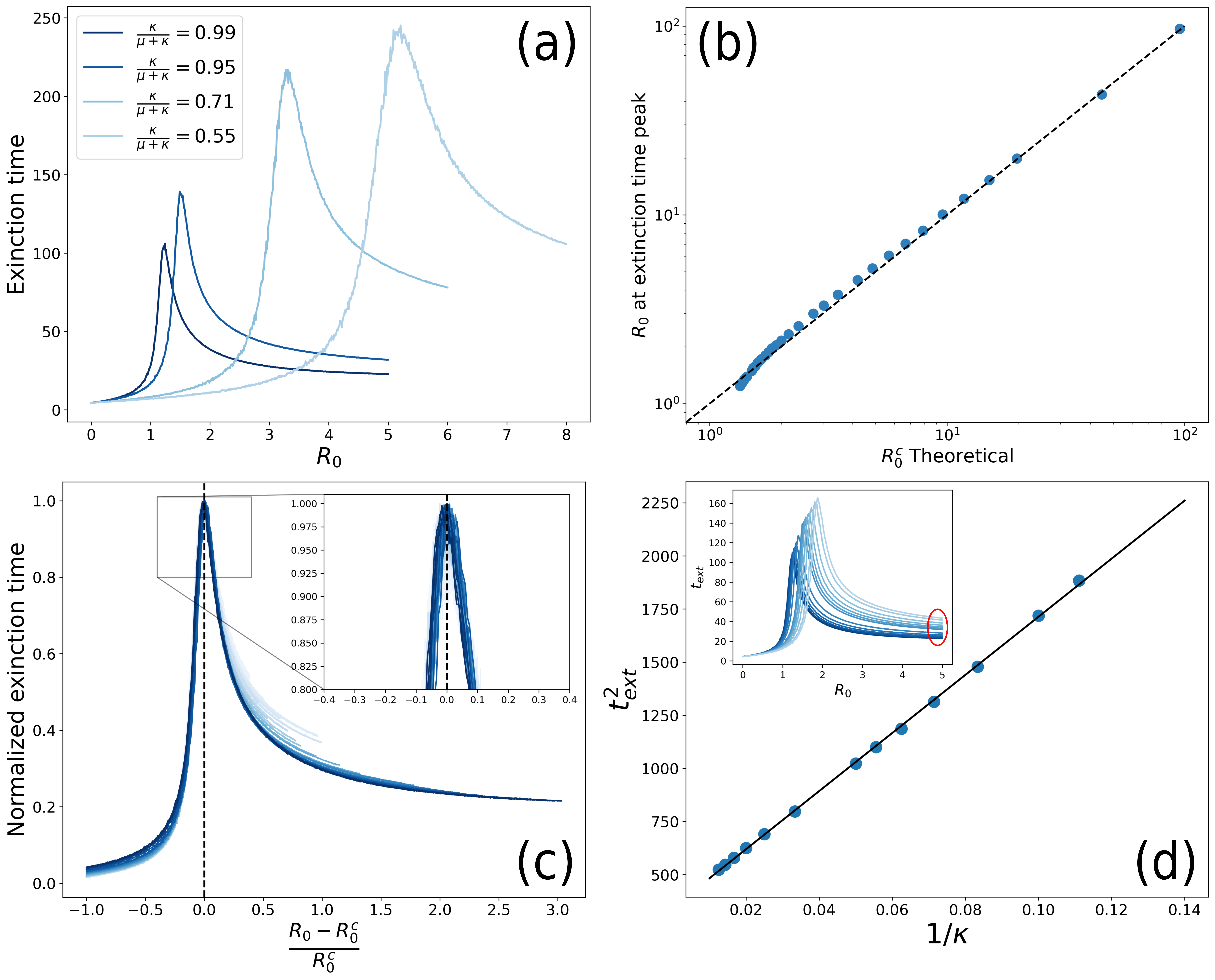}
        \caption{(a) Extinction time for some values of the parasite mobility. (b) Comparison of the critical $R_0$ value computed with \cref{eq:R_0ibm} compared to the values obtained numerically by computing the maximum of the extinction time. (c) Scaling of the extinction time with several values of the parasite mobility. (d) Representation of the square of the extinction time as function of the inverse of the parasite mobility. The inset shows the zone where this relation has been computed, showing a good agreement with \cref{eq:t_ext}.}
        \label{fig:extinction_time}
    \end{figure*}
    
    The extinction time is defined as the time elapsed from the beginning of the epidemic until the system reaches its absorbing state, that is, when no parasites or infected individuals are left. From \cref{eq:front_vel}, we expect the time to extinction to increase when the parasite mobility is decreased. Moreover, we expect the extinction time to decrease with the distance to the epidemic threshold, as we expect to reach faster the absorbing state for larger values of the spatial basic reproduction number. 
    
    In the limiting case where all (or almost all) hosts die, it is clear that the disease must have spread to the entire system. Thus, in this limit, the extinction time should be proportional to the inverse of the disease spreading speed, $t_{\textrm{ext}}\sim 1/v$. Then, in this limit, we can relate the extinction time with the parasite mobility as follows,
    \begin{equation}\label{eq:t_ext}
        t_{\textrm{ext}}\sim\frac{1}{\sqrt{\kappa}} \quad \textrm{for} \quad \tilde{R}_0\gg1 \ .
    \end{equation}
    However, the absorbing state is not always reached after all hosts becoming infected and \cref{eq:t_ext} is only expected to work far from the epidemic threshold, when the disease is expected to spread to the entire system.
    
    In \cref{fig:extinction_time}(a) the extinction time is plotted against the basic reproduction number for some values of the parasite mobility. As expected, the extinction time increases for lower values of the parasite mobility. The increasing behaviour before the peak can be understood as the increasing time needed for the initial perturbations to decay to the disease-free phase. After the peak, the greater the basic reproduction number the faster the epidemic will reach its absorbing state with a non-negligible number of dead individuals. So, with this interpretation, the peaks of the extinction time should coincide with the epidemic threshold for each value of the parasite mobility.
    
    In \cref{fig:extinction_time}(b) we compare the numerical value of $R_0$ at which the extinction time peaks with the theoretical value of $R_0^c$, computed with \cref{eq: threshold_corrections}, showing good agreement. Thus, the dependence of the extinction time with $R_0$ should vanish if plotted against the distance to $R_0^c$. Furthermore, if the extinction time is normalised (dividing each line by its maximum), all the lines should collapse near the transition point. In \cref{fig:extinction_time}(c) the normalised extinction time is plotted against the distance to the critical value of $R_0$. The scaling is shown to be valid only near the transition point, as expected.
    
    In the limiting case where the epidemic dies by infecting a large part of the host population, i.e. for a large enough $R_0$ value, the extinction time should follow \cref{eq:t_ext}, as previously discussed. In \cref{fig:extinction_time}(d) we show how the extinction time relates to the parasite mobility in this limit, following the predicted behaviour.
    
    \section{Conclusions} \label{sec: conclusions}

    In this work we have developed a spatially-explicit individual-based model for parasite-produced marine epidemics of immobile hosts. This study has allowed us to tackle important questions in marine epidemiology, as how spatial constraints affect epidemic spreading in filter-feeder populations or how will the infected population of hosts change in space and time. While addressing the aforementioned questions, we have shown that there exists a regime of high parasite mobility where the time progression of both host and parasite populations can be well described by the non-spatial version of the model (i.e. the system of ODE's presented in \cite{GimenezRomero2021}). We have also shown that a fast-slow approximation for the time progression of the parasite population, already presented in \cite{GimenezRomero2021}, can be extended for spatial systems. Interestingly, the conditions under which this approximation is valid  are less restrictive than in the non-spatial case, and regimes in which this approximation is valid for low mobility and comparable time scales are reported in this contribution.
    
    We have derived an approximate analytical expression of the \textit{spatial} basic reproduction number, that allows to predict the onset of a global epidemic in a spatial model. The obtained expression explicitly shows a trade-off between the intrinsic pathosystem dynamics (i.e. $R_0$) and a factor accounting for parasite mobility. Moreover, the spatial threshold defined by $\tilde{R}_0=1$ separates the final state of the system in two different phases, namely a disease-free phase and a propagation phase. In the propagation phase, any initial condition of infected individuals or parasites will propagate throughout the system, causing a proper outbreak. On the other hand, in the disease-free phase the conditions are not sufficient for a local introduction of parasites or infected individuals to spread through the system. The effect of the parasite mobility in the spatial basic reproduction number is clear, the more parasites move the more infections they cause.
    
    The spatiotemporal behaviour of the system has been investigated in the propagation phase. First, we showed that the infected population spreads through the space with a speed directly proportional to the square root of the diffusion coefficient of parasites, showing good agreement between the derived analytical expression and numerical simulations. The time to extinction has been also studied by means of numerical simulations, showing that, if the system is far above enough of the spatial threshold, the time to extinction can be analytically computed, in good agreement with simulations. We obtained that larger values of the parasite mobility yield more severe epidemics in which there are more infections and the extinction is faster.
    
    To summarise, in the present work we have introduced and analysed and Individual-Based approach to epidemic transmission in spatially extended systems of immobile hosts. The infection mechanism is due to mobile parasites, that are in turn produced by infected hosts. The study allows to answer some biologically relevant questions, like predicting the occurrence of a global epidemic outbreak or its velocity of expansion through the system. Thus, the analytical and computational results of the model shed light on the underlying mechanisms underpinning the emergence of a global epidemic outbreak and its spatial progression. This work provides a first step into the spatial-explicit individual-based modelling of marine epidemics of immobile hosts. 
    
    Although this work has considered the case of a spatially homogeneous distribution of hosts, we plan to extend the study to more general cases, discussing the effect of inhomogeneous spatial host distributions. Furthermore, other biological relevant effects could be added to the model to enhance the description of different epidemics, e.g. infected individuals could still filter parasites or parasite-load dependent infection process. The model could also describe epidemics on other immobile species such as filter feeders like sponges or other bivalves, corals, intertidal communities or starfishes provided that the necessary modifications in the model are properly included. Stochastic spatially-explicit descriptions like the one presented here could be also extended to the study of epidemics of other immobile hosts, like vector-borne diseases of plants. However, this would imply a quite different model to describe the different epidemic compartments of the vectors and also their ecological features. We hope these studies can be useful in conservation plans or ecosystem management and could serve as a basis for more sophisticated models.
    
\appendix

\section{Derivation of the non-spatial equation for $R_{\infty}$} \label{app:Rinf}

    The model described by the ODE system in \cref{eq:SIRP_MF} has a conserved quantity $\mathcal{C}$ given by \cite{GimenezRomero2021}.
    \begin{equation}\label{eq:conservedquantity}
        \mathcal{C}=P + \frac{\lambda}{\gamma}\parentesi{S+I}-S-\frac{\mu}{\bar{\beta}}\ln S
    \end{equation}
    
    At $t=\infty$ the system reaches an absorbing state completely determined by $S(\infty)$, as $P(\infty)=I(\infty)=0$ and $N=S(\infty)+R(\infty)$. Thus, from \cref{eq:conservedquantity} we have
    \begin{equation}\label{eq:transcendental}
        S(\infty)\parentesi{\frac{\lambda}{\gamma}-1}-\frac{\mu}{\beta}\ln(S(\infty))=\mathcal{C}_0
    \end{equation}
    
    The transcendental equation \cref{eq:transcendental} can be solved by means of the Lambert's W function,
    \begin{equation}
        S(\infty)=-\frac{\mu\gamma}{\beta\parentesi{\lambda-\gamma}}W_0\parentesi{-\frac{\beta\parentesi{\lambda-\gamma}}{\mu\gamma}\exp(-\beta\mathcal{C}_0/\mu)}
    \end{equation}
    which can be simplified to
    \begin{equation}\label{eq: S_inf}
        S(\infty)=-\frac{S(0)}{\xi}W_0\parentesi{-\xi\exp(-\frac{\beta}{\mu}C)} \ ,
    \end{equation}
    with $\xi=S(0)\frac{\beta\parentesi{\lambda-\gamma}}{\mu\gamma}$ and $C=P(0)+\frac{\lambda}{\gamma}\parentesi{S(0)+I(0)}-S(0)$.\\
    
    Finally, the absorbing state fulfils the condition $N=S(\infty)+R(\infty)$ so that the final number of dead individuals can be expressed as
    \begin{equation}
        R(\infty)=N+\frac{S(0)}{\xi}W_0\parentesi{-\xi\exp(-\frac{\beta}{\mu}C)} \ .
    \end{equation}
    
\textbf{Data Accessibility}
The computer codes used in the numerical simulations can be found in the GitHub repository \cite{CODE}. \\
    
\textbf{Authors' contributions.} AGR and MAM designed research. AGR wrote the computational code with input from FV in the IBM implementation. AGR performed the numerical simulations. AGR, FV and MAM carried out analytical derivations. AGR, FV, CL and MAM analysed and discussed the results. AGR and MAM wrote the paper. All authors read and approved the final version of the paper.\\

\textbf{Competing Interests.} The authors declare that they have no conflict of interest.\\
    
\textbf{Funding.} Grant RTI2018-095441-B-C22 (SuMaEco) funded by MCIN/AEI/10.13039/501100011033 and by “ERDF A way of making Europe" (AGR and MAM). Grant MDM-2017-0711 (María de Maeztu Excellence Unit) funded by MCIN/AEI/10.13039/501100011033 (AGR, CL and MAM). FV  through the \textit{Professors Convidats} Program funded by UIB.


%

\end{document}